# Revealing the Phonon Bottleneck Limit in Negatively Charged CdS Quantum Dots

*Skylar J. Sherman[1], Bokang Hou[2,3], Matthew J. Coley-O'Rourke[2], Katherine E. Shulenberger[1,†], Lauren M. Pellows[1], Eran Rabani[2,3,4], Gordana Dukovic[1,5,6*]*

1. Department of Chemistry, University of Colorado Boulder, Boulder, Colorado 80309, United States

2. Department of Chemistry, University of California Berkeley, Berkeley, California 94720, United States

3. Materials Sciences Division, Lawrence Berkeley National Laboratory, Berkeley, California 94720, USA

4. The Raymond and Beverly Sackler Center of Computational Molecular and Materials Science, Tel Aviv University, Tel Aviv 69978, Israel

5. Department of Chemistry and Renewable and Sustainable Energy Institute (RASEI), University of Colorado Boulder, Boulder, Colorado 80309, United States

6. Materials Science and Engineering, University of Colorado Boulder, Boulder, Colorado 80303, United States

Abstract:

The capture of photoexcited hot electrons in semiconductors before they lose their excess energy to cooling is a long-standing goal in photon energy conversion. Semiconductor nanocrystals have large electron energy spacings that are expected to slow down electron relaxation by phonon emission, but hot electrons in photoexcited nanocrystals nevertheless cool rapidly by energy transfer to holes. This makes the intrinsic phonon-bottleneck limited electron lifetime in nanocrystals elusive. We used a combination of theory and experiments to probe the hot electron dynamics of negatively charged Cadmium Sulfide (CdS) colloidal quantum dots (QDs) in the absence of holes. Experiments found that these hot electrons cooled on a 100 ps timescale. Theoretical simulations predicted that pure phonon-bottleneck limited electron cooling occurs on a similar timescale. This similarity suggests that the experimental measurements reflect the upper limit on hot electron lifetimes in these CdS QDs and the lower limit on the rates of processes that can harvest those hot electrons.

## Introduction:

The research field of semiconductor nanocrystals has long been captivated by the quest for long-lived hot carriers. In bulk semiconductors, photoexcited hot carriers cool to the band edges so rapidly compared to other relaxation processes that, for many practical purposes, the excess energy of the hot carrier is considered lost to lattice heating.[1] In nanocrystalline semiconductors, the electronic states near the band edges have discrete, rather than continuous, energy spacings that are large compared to phonon energies.[2] These large energy spacings have been predicted to cause a "phonon bottleneck" that would cause significantly slower carrier cooling in nanocrystals compared to their parent bulk semiconductors.[3-6] Slow carrier cooling would, in turn, enable efficient hot carrier extraction in a photovoltaic device or a photochemical system, capturing a greater fraction of the absorbed photon energy.[7]

In practice, however, electrons in nanocrystals usually cool on a sub-picosecond timescale, which has been explained with other cooling mechanisms in this size regime, including coupling to surface traps and ligand vibrations, as well as energy transfer from hot electrons to generate hot holes.[8] The latter is thought to be the dominant mechanism of electron cooling that renders the much slower phonon-coupled relaxation essentially irrelevant.[1, 9, 10] Accordingly, strategies for slowing down hot electron cooling in nanocrystals have involved spatially decoupling the electron and hole via band-alignment engineering, surface functionalization with hole-localizing ligands, and filling of the photoexcited hole by chemical reduction or electrochemical reduction.[11-20] In CdSe-based nanocrystals, these approaches have extended hot electron lifetimes to picosecond and even nanosecond timescales, illustrating that it is possible to manipulate electron cooling channels to slow down cooling rates.[13, 15-20] However, such strategies can simultaneously change the energy level spacing, surface-capping ligands, and the trapping landscape, making it difficult to disentangle the contributions from the different cooling mechanisms to the observed cooling rates. Due to this complexity, the true phonon bottleneck limit, i.e., the upper limit for the hot electron lifetime where the cooling rate is determined by lattice phonon-assisted relaxation, has remained elusive. Consequently, the lower limit for the rate of an electron harvesting process that could be competitive with cooling has not been determined.

In this work, we use a combination of theory and experiments to establish the lower limit for electron cooling rate using negatively charged CdS quantum dots (QDs). In these particles, the energy transfer from hot electrons to holes is removed as a cooling mechanism. Experimentally, we generated negatively charged ~4.5 nm diameter CdS QDs (QD$^{(-)}$) *in situ* in a transient absorption (TA) spectrometer by illuminating with a continuous wave laser, without added reducing agents. When a QD$^{(-)}$ absorbs a photon, a negative trion is generated. A subsequent nonradiative process known as Auger or Auger-Meitner recombination annihilates one electron and one hole and generates a hot electron (Scheme 1). We probed the relaxation of those hot electrons by TA spectroscopy and identified spectroscopic signatures of the exciton, trion, and hot electron species. We found that the hot electrons generated via trion recombination cooled to the conduction band edge with a time constant of ~100 ps. We attribute this cooling timescale to the relaxation of the hot electron from the 1P state to the band edge (i.e., the 1S state). This relaxation would require coupling to lattice or ligand vibrations, relaxation via trap states, or a combination of these mechanisms.

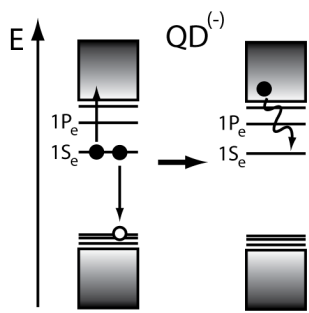

**Scheme 1.** Generation of a hot electron in a QD$^{(-)}$ by nonradiative trion recombination and the subsequent cooling of the electron by intraband relaxation.

To elucidate the mechanism of electron cooling in these CdS QD$^{(-)}$, we examined the relaxation from the manifold of 1P states to the 1S state via coupling to lattice vibrations. We predict the associated cooling dynamics by parametrizing a model Hamiltonian with electron energies and electron-phonon couplings computed using an atomistic semiempirical pseudopotential model,[21, 22] which included long-range Fröhlich-type couplings.[23] These simulations revealed a lifetime of $2 - 144$ ps for QDs in the 4-5 nm diameter range. The similarity between the calculated and measured electron lifetimes suggests that internal lattice vibrations and multiphonon processes with sufficiently strong coupling to the optical phonons are sufficient to account for the relaxation of hot electrons in CdS QD$^{(-)}$, without the need to invoke coupling to ligand vibrations or trap states. While these other relaxation channels cannot be ruled out, it appears that the relaxation of hot electrons in these CdS QD$^{(-)}$ can be described by a pure lattice-phonon bottleneck, resulting in slow relaxation that is consistent with recent promising photocatalysis results with CdS QDs.[24, 25]

**Results:**

***In Situ* TA Spectroscopy of Photoreduced CdS QDs**

The CdS QDs studied in this manuscript have and are capped with oleate surface-capping ligands. The first excitonic transition in the absorption spectrum of these QDs (Figure 1a) has a peak at 443 nm. Using this wavelength, the diameter is predicted to be between 3.93 nm[26] and 5.02 nm[27], depending on the sizing curve used. Due to this uncertainty, we estimate our QDs to have a 4.5 nm diameter. When illuminated with above-bandgap photons in an inert atmosphere, these QDs build up a population of $QD^{(-)}$ on the timescale of 10-20 minutes until an equilibrium between charged and neutral QDs is reached.[28] When illumination is halted, the QDs return to neutral on the order of minutes. Figure 1a shows an example of how the absorbance of the QDs changes as a function of 405 nm continuous wave (CW) laser power that hits the sample about 1 cm below the absorbance collection point. Figure 1b shows the assignments of the peaks in the absorption spectra.[2, 29, 30] With increasing CW power, we observe a reduction to the 1S and 2S transition strengths, along with a red shift of the 1S, 2S and 1P transitions, both attributed to the presence of an electron at the conduction band edge.[31-35] Previously, we hypothesized that the photo-reduction mechanism involves the oxidation of an oleate ligand by the photoexcited hole and subsequent dissociation of the oxidized ligand, leaving the QD negatively charged.[28] This mechanism is distinct from other reported photoreduction reactions of QDs that involve hole scavenging by chemical reducing agents,[18, 33-38] and avoids chemical complications from additional species in solution.

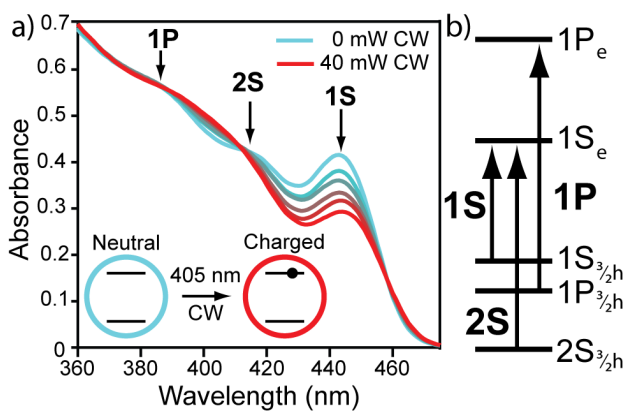

**Figure 1.** (a) UV-Vis absorption spectra of CdS QDs after irradiation with 0, 5, 10, 20, 30 and 40 mW of 405 nm CW laser power. (b) Energy level diagram showing the relevant states for electrons and holes and the absorption transitions between them.

To experimentally probe the excited state dynamics of the CdS QD$^{(-)}$, we modified a TA setup to include a 405 nm CW laser, which illuminated the sample approximately 1 cm away from the point at which the pump and probe overlap (Figure 2a). The degree of photocharging was controlled by varying the power of the 405 nm CW laser. The short timescale dynamics in neutral and negatively charged CdS QDs after pump excitation are described in Section S1 and shown in Figure S1. Briefly, initial electron cooling after excitation of neutral CdS QDs occurs on a 50 fs timescale, similar to a previous report,[39] and consistent with fast cooling observed in QD systems when the hole is present.[1, 9, 10] In photoexcited QD$^{(-)}$ the electrons cool with a time constant of 5 ps due to a previously described spin blockade effect.[18] To study the subsequent relaxation dynamics, we focus on timescales after 30 ps, which is late enough that processes like initial carrier cooling and hole trapping are complete but is early enough that exciton and trion decay are not yet significant.

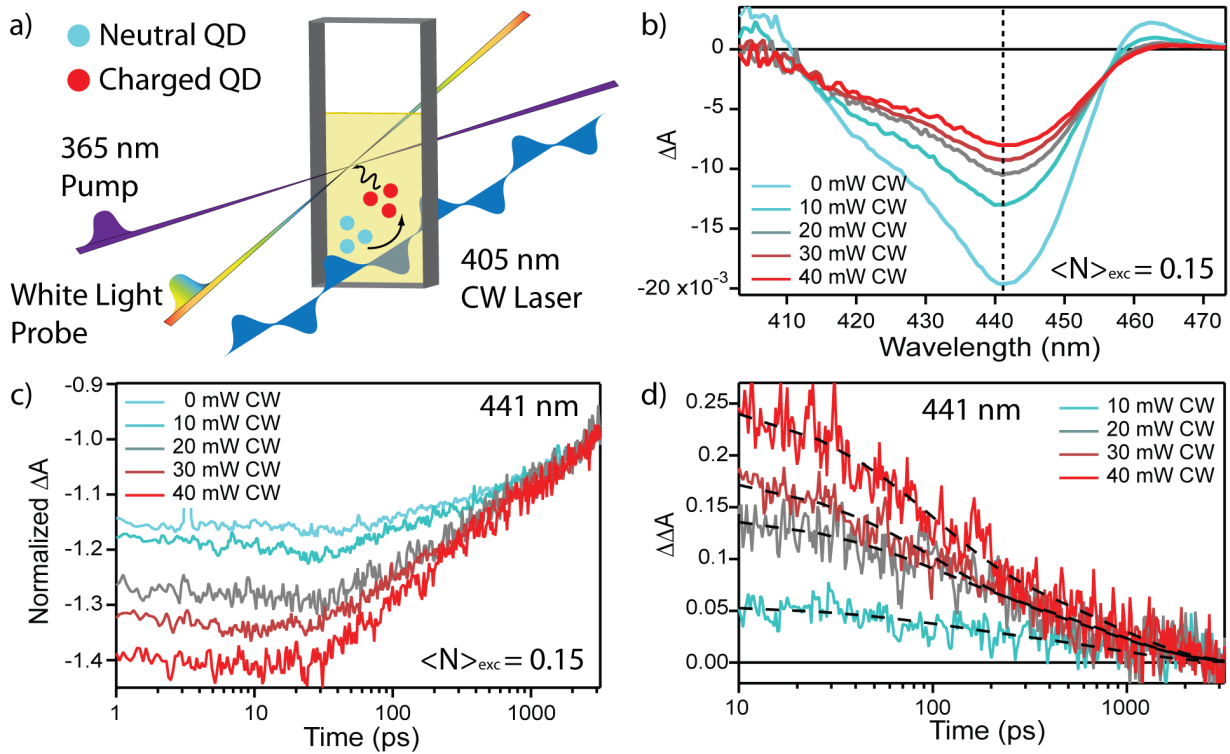

**Figure 2.** (a) Schematic of the TA experiment. The pump and probe remain in the standard TA alignment while the CW beam illuminates the cuvette ~1 cm from the spot where the pump and probe overlap. This beam controls the equilibrium between charged (red spheres) and neutral QDs (teal spheres). (b) ΔA spectra at 30 ps for a range of CW powers. (c) Time-traces at 441 nm, which corresponds to the bleach maximum shown as a dashed line in (b), for a range of CW powers. (d) Difference TA traces where the 441 nm trace for the un-illuminated (0 mW) sample is subtracted from the signal of illuminated samples shown in (c) along with a fit to a biexponential.

Figure 2b shows TA spectra collected at a pump probe delay of 30 ps for 5 CW powers: 0, 10, 20, 30, and 40 mW. The pump power was kept low to avoid complications from higher order excitations (Section S2 and Figure S2). At 0 mW we observe features corresponding to excitons that have been thoroughly discussed in the literature.[40] Briefly, the exciton ΔA spectrum is composed of a bleach of the 1S (441 nm) and 2S (419 nm) exciton absorbances and a positive absorbance (PA) on the red side of the 1S bleach (>457 nm). The 1S and 2S bleaches are due to state filling by photoexcited electrons at the conduction band edge.[40] The PA on the red edge is due to bandgap renormalization which shifts the bandgap to a lower energy and creates a new transition with lower energy than the ground state 1S absorbance.[40] With increasing CW power, and increasing population of QD$^{(-)}$, we see a loss of the PA feature because negative trions have a fully occupied conduction band edge, and the strength of the 1S and 2S bleaches decreases. We show these changes occurring in real time during charging and discharging in Figure S3.

Figure 2c shows ΔA time-traces at the bleach maximum of 441 nm for the five CW powers. Because trions decay much faster than excitons,[18, 28, 36-38] the traces are normalized at 2.5 ns, and we see increased contributions from the relatively faster trion decay with increasing CW power (and thus more charged QDs). To isolate the trion-only dynamics, we subtract the neutral trace (0 mW CW power) from the charged traces when normalized at late times (ΔΔA, Figure 2d).[18, 36, 41, 42] If the observed dynamics for charged QDs reflected only trion recombination, the difference between neutral and charged traces would yield a single exponential decay with the rate being equal to the rate of trion decay.[18, 36, 41, 42] The difference traces in Figure 2d require a biexponential to fit the data. We globally fit the four traces shown in Figure 2d to a biexponential decay with the rate constants linked. The longer time component of the biexponential yields a time constant of 826 ps, which is similar to a previously found trion nonradiative recombination time constant of 1.3 ns in 4.8 nm diameter CdS QDs.[36] The faster component, with a time constant of 90 ps, indicates the presence of another species, besides trions, contributing to the observed ΔΔA. We hypothesize this species to be hot electrons generated by nonradiative trion recombination. Further evidence for this species is described later in the text.

**Theoretical Simulation of Hot Electron Cooling**

To evaluate the hypothesis that the ~100 ps timescale is associated with hot electron cooling in QD$^{(-)}$, we carried out computer simulations of the nonradiative relaxation of an excited electron, where initially the electron occupies the highest $1P_e$ electronic level (see Figure 3(a)) and the lattice phonons of the NC assume a canonical thermal distribution. We simulated the relaxation of this excess electron for 4.5 nm CdS QDs using a model Hamiltonian parameterized by a semiempirical pseudopotential model (see Sections S4 and S5 for more details).[21, 43, 44] Due to negligible overlap with ligands, we ignored the coupling to the ligand modes.[23] The relaxation dynamics were described by a non-Markovian quantum master equation, with rates determined by Fermi's golden rule.[45, 46] In Figure 3(b) we plot the $1S_e$, $1P_{e1}$ and $1P_{e3}$ states, which exhibit similar symmetry to the states predicted by the effective mass model. The energy gap between the $1S_e$ and $1P_e$ levels calculated from the pseudopotential model is $\Delta\varepsilon \approx 180$ meV. In Figure 3(c) we plot the population dynamics of all four states involved. The 1P$_e$ states mix on ultrafast timescales (hence the rapid drop in their respective populations), followed by a slower exponential decay to the 1S$_e$ state. The rate calculated from a fit to the decay of the populations is $k \approx 43$ ns$^{-1}$ ($\tau \approx 23$ ps). Since the relaxation rate scales exponentially with the energy gap, as discussed in Section S6, the uncertainty in the experimental determination of QD diameter leads to a significant spread in the theoretical values that may correspond to the experimental sample. Additionally, the experimental sample is not perfectly monodisperse so there is a distribution of experimental diameters and corresponding relaxation rates that is contained in the experimental average. For QD diameter range of 4.0 – 5.0 nm (the average diameter of the QDs in the experimental sample is inside this range), the calculated time constants span 144 – 2 ps, respectively. The experimentally measured rate constant falls inside this range.

The inset in Figure 3(c) shows the spectral density characterizing the strength of electron-nuclear couplings. The major contribution to the spectral density and hence, to the relaxation of the hot electron, comes from three frequency regimes. The lowest frequency is associated with a surface mode and the other two are optical modes, as shown in Figure 3(d). The strongest coupling occurs with an out-of-phase optical mode that involves the interior Cd and S atoms, as anticipated for long-range electronic coupling to longitudinal optical (LO) phonons by Fröhlich interactions.[47, 48]

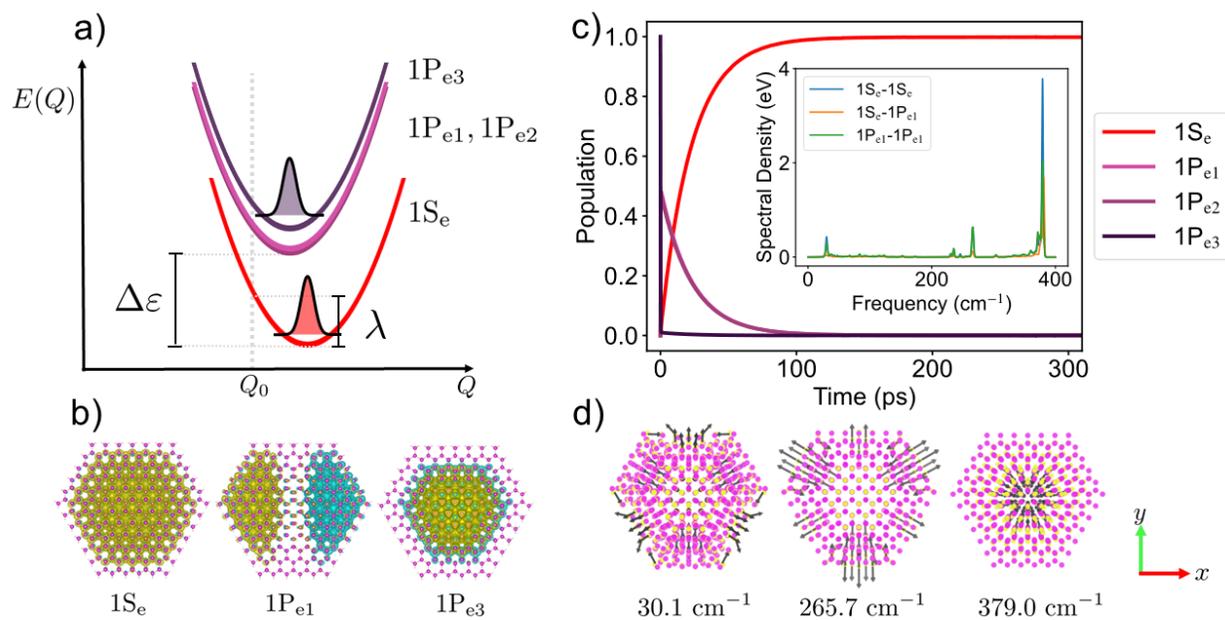

**Figure 3.** (a) Schematic picture of the diabatic surfaces for $1S_e$ and $1P_e$ states. $\Delta\varepsilon \approx 180$ meV is the energy gap between $1P_e$ and $1S_e$ states and $\lambda \approx 40$ meV is the reorganization energy for $1S_e$ state. (b) Isosurface plots for $1S_e$ and $1P_e$ wavefunctions at the equilibrium configuration $Q_0$. The yellow and blue show positive and negative phases. (c) Calculated relaxation kinetics of the excess electron for a 4.5 nm CdS QD. The populations for $1P_{e1}$ and $1P_{e2}$ are overlapped due to energy degeneracy. Inset: Spectral densities for intrastate ($1S_e$-$1S_e$ and $1P_e$-$1P_e$) and interstate couplings ($1S_e$-$1P_e$). (d) Phonon vibrations with major contributions to the reorganization energy and the relaxation dynamics. Videos of these phonon modes are provided in Supporting Videos 1-3.

**Extraction of Hot Electron Spectroscopic Signature and Cooling Rate from TA Data**

The computational results support the hypothesis that the ~100 ps relaxation timescale observed experimentally corresponds to electron cooling in CdS QD$^{(-)}$. To further test this hypothesis, we go beyond single wavelength kinetics and look for spectral signatures of the relevant species and their associated decay constants. We expect that each relevant species has a unique ΔA spectrum that decays with the kinetics associated with that species,[49-52] resulting in a total ΔA spectrum that is a population-weighted sum of these individual spectra. We first describe a kinetic model for the relevant processes in the 30 ps – 3 ns time window and predict the approximate form of the ΔA spectra of the species we expect to observe. We then use target analysis methods to extract decay-associated spectra from experimental data, compare them to the predicted spectra to support their assignments, and determine the rate constants of exciton, trion, and hot electron decay.

Photon absorption in a neutral QD creates an exciton that will decay to the ground state with a rate constant of $k_x$ (Figure 4a). Photon absorption in a QD$^{(-)}$ creates a negative trion (Figure 4b). A negative trion will decay nonradiatively with a rate constant $k_t$, transferring the energy to the remaining electron making it hot. We ignore the possibility of radiative trion decay as it is predicted to be slower than nonradiative decay in these core-only QDs.[53, 54] This hot electron then cools back to the QD$^{(-)}$ ground state with a rate constant $k_h$. This leads to a coupled set of rate equations that describe the populations of each state with time as described in Section S7 (Equations S31-S33).

Next, we first predict the approximate spectral signatures of these species. We fit the steady-state absorbance for the neutral QDs (Figure 1a) to three gaussians to find the position, amplitude, and width of the 1S, 2S, and 1P transitions. Fits are completed with an energy x-axis and converted back to a wavelength x-axis for comparisons. The result of this fit is shown in Figure 5c as the neutral trace. For a photoexcited neutral QD after carrier cooling, the TA spectrum is the sum of a bleach of transitions involving the conduction band edge and a renormalization of the bandgap.[40] Because these are core-only QDs which have a low photoluminescence quantum yield, stimulated emission does not contribute significantly to the observed ΔA spectra.[55-57] Following this logic, we simulate the excited state absorbance spectrum for a neutral QD by reducing the 1S and 2S amplitudes by 60% (chosen based on literature precedent[50, 51, 58]) and shifting their transition energies by 55 meV (chosen so that the predicted ΔA spectrum would match the relative peak positions and intensities of the 1S bleach and the PA feature in the experimentally measured TA spectrum of neutral QDs shown in Figure 2b), leading to the exciton absorbance spectrum shown in Figure 4c. The corresponding ΔA spectrum for the exciton is determined by subtracting the steady-state absorbance spectrum for the neutral QD from the absorbance spectrum for the exciton and is shown in Figure 4d.

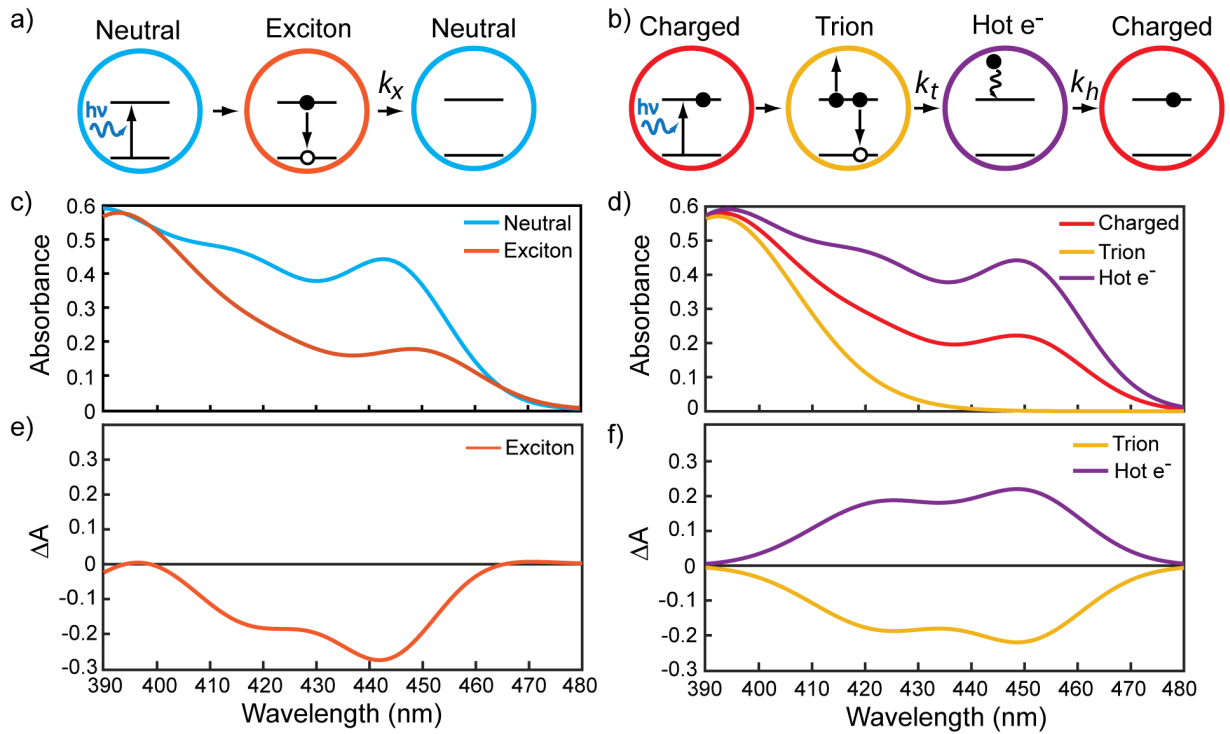

**Figure 4.** (a) Schematic of exciton generation and decay in a neutral QD. (b) Schematic of trion generation followed by nonradiative decay to create a hot electron which decays back to the conduction band edge in QD$^{(-)}$. (c) Fit steady-state absorbance spectrum for neutral QDs (blue) and predicted absorbance spectrum of an exciton (orange). (d) Predicted ΔA spectrum for an exciton. (e) Predicted steady-state absorbance spectrum for negatively charged QDs (red) and predicted absorbance spectra for a trion (yellow) and hot electron (purple). (f) Predicted ΔA spectra for a trion (yellow) and hot electron (purple).

Next, we predict the ΔA spectra of trions and hot electrons in QD$^{(-)}$'s. Because the hole contribution to the band gap bleach is minimal in core-only Cd-chalcogenide particles,[40, 59, 60] the change in absorbance upon charging is very similar to the change due to excitation, with a reduction the absorbance strength of the 1S and 2S transitions and a shift of the bandgap energy.[18, 25, 28, 33, 36-38, 61] For the purposes of predicting the ΔA spectral shapes of trion and hot electrons we use the assumption that the 1S and 2S transitions are bleached by 50% when an electron occupies the conduction band edge.[18, 28, 33, 36-38, 61] We predict the excited state absorbance spectrum of a cold trion by fully bleaching the 1S and 2S transitions (because the conduction band edge is doubly occupied), leaving only the 1P absorption (Figure 4e). We predict the hot electron excited state spectrum by setting the 1S and 2S transitions to the same amplitude as the neutral QD absorbance (because there is no electron at the conduction band edge) while shifting the band gap by 55 meV (consistent with the band gap shift due to charging seen in Figure 1a). We then predict the ΔA spectra for trions and hot electrons by subtracting the charged ground state absorbance spectrum from the absorbance spectra of the trion and hot electron (Figure 4f). The hot electron TA spectrum has a positive ΔA because the lack of occupation of electrons at the conduction band edge leads to a recovery in absorbance at the bandgap energy and thus a positive ΔA signature.

Finally, we fit the observed TA data using a target analysis methodology.[49-52] To accomplish this, we assume that the exciton, trion and hot electron states have an associated unique ΔA spectrum: $\Delta A_x$, $\Delta A_t$, and $\Delta A_h$, respectively. We define the initial populations of excitons and trions generated by the pump as $N_{x0}$ and $N_{t0}$ respectively. The populations of each state are governed by the first order kinetics process schematically shown in Figure 4a and 4b and explicitly defined in Section S7 (Equations S31-33). As detailed in Section 7, the total TA spectrum evolves over time according to Equation 1:

$$\Delta A(\lambda, t) = \Delta A_x(\lambda) N_{x0} e^{-k_x t} + N_{t0} e^{-k_t t} \left( \Delta A_t(\lambda) + \Delta A_h(\lambda) \frac{k_t}{k_h - k_t} \right)$$
$$- \Delta A_h(\lambda) \frac{k_t}{k_h - k_t} e^{-k_h t} \quad (1)$$

To extract $\Delta A_x$, $\Delta A_t$, and $\Delta A_h$, we globally fit the entire TA data set (10-40 mW CW power, spectral range 403 to 473 nm, 278 time points from 30 ps to 3 ns) to equation 1. We start the fit at 30 ps to avoid contributions from the spin blockade which would affect the dynamics at early times.[18, 62] We require the rate constants $k_x$, $k_t$ and $k_h$ to be equal at all wavelengths and CW powers. $N_{x0}$ and $N_{t0}$ are held equal for each wavelength at a CW power. The comparison of fit to TA data is shown in Figure S5 and the $\Delta A$ component spectra for each power are shown in Figure S6.

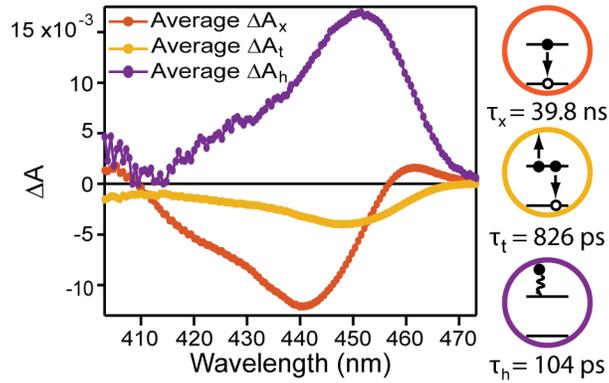

**Figure 5.** Spectral components for the exciton ($\Delta A_x$), trion ($\Delta A_t$), and hot electron ($\Delta A_h$) extracted from a global fit to equation 1 averaged for the 4 powers. Schematics of each state along with their corresponding decay time constant shown to the right.

The resulting averaged spectral components associated with time constants of 39.8 ± 1.2 ns, 826 ± 0.25 ps, and 104 ± 1.7 ps are shown in Figure 4c. The fitted values for $N_{x0}$ decrease and $N_{t0}$ increase with increasing CW fluence and are presented in Table S3. We do not interpret the time constant for exciton decay, as there is minimal exciton decay in the time window studied. Based on the resemblance to the predicted spectra in Figure 4 and the independent insights into the timescales (the reported rate of trion decay[36] and the rate of electron cooling theoretically simulated above), we can assign the 826 ps and 104 ps components with a high degree of confidence. The 826 ps component is similar in timescale to the previously measured trion decay rate in CdS QDs of a similar size,[36] and the associated spectrum, specifically the lack of PA on the red side of the bleach along with a redshift of the bleach maximum, matches well the predicted TA spectrum of trions (Figure 4). We discuss the trion bleach intensity in Section S8. For the 104 ps component, the associated spectrum resembles the predicted TA spectrum of hot electrons (Figure 4), namely it exhibits positive features at the 1S and 2S transition energies. Its decay rate is also remarkably close to the theoretically predicted hot electron cooling rate in the absence of the hole (Figure 3). We therefore assign the ~100 ps process to cooling of hot electrons generated by nonradiative trion recombination in CdS QD$^{(-)}$. While trion recombination is expected to generate very hot electrons, the slowest step during cooling is from the 1P to the 1S electron states because the density of electron states increases significantly further away from the band edge.[23] Hence, our measurement reports approximately the rate constant for 1P to 1S relaxation.

Our experimental results and analysis are quite robust. To test whether the cooling dynamics change upon prolonged illumination, we repeated the TA experiments and analysis on the same batch of CdS QDs that been illuminated with the 405 nm CW laser at an average power of 13.4 mW for 8.3 hours. Figure S7 shows the resulting component spectra which are very similar to the ones in Figure 5. The time constants for decay are found to be 46 ns, 860 ps, and 109 ps for the exciton, trion, and hot electron, respectively. These values are very similar to those reported in Figure 5.

To test for sample-to-sample variation, we conducted the TA experiments and analysis on another batch of QDs from a different synthesis with an estimated average diameter of 4.6 nm (Figure S8a). Upon illumination with the 405 nm CW laser, we observe similar changes to the spectrum at 20 ps (Figure S8b) and kinetics at the band edge bleach (Figure S8c). Using the same global fitting procedure as above, we obtained a trion decay constant of 827 ps and hot electron cooling time constant of 129 ps with similar spectral signatures for the exciton, trion and hot electron (Figure S8d). These time constants are quite close to the time constants reported in Figure 5.

## Discussion:

The agreement between the theoretically calculated range and experimentally measured electron cooling rate constant leads us to the conclusion that hot electron relaxation in $QD^{(-)}$ can originate from the Fröhlich coupling to optical modes resulting in slow multiphonon emission, reaching the phonon bottleneck limit. While coupling to ligand vibrations or transit through trap states could also facilitate hot electron cooling, it seems unnecessary to explain the experimentally observed timescales. Furthermore, the similarity in cooling rates measured when QDs are irradiated for extended periods or a different batch of QDs with similar diameter is studied, supports the notion that the cooling kinetics are governed by an intrinsic property of the QD lattice and not surface properties that vary between samples. Consequently, we conclude that we have observed the phonon-bottleneck limit for hot electron cooling in CdS QD, i.e., the hot electron lifetime achievable when other cooling channels are removed. This means that, in the phonon bottleneck regime, electron transfer processes with rate constants above $10^{10}$ s$^{-1}$ can outcompete cooling to harvest hot electrons. There are multiple electron acceptors that accept electrons from QDs with rate constants fast enough that they could compete with the cooling shown here.[63, 64]

To place the phonon bottleneck timescale for these CdS QDs in context, the timescale is similar in magnitude to the 750 ps measured for epitaxially grown InGaAs quantum dots at 40K.[14] Furthermore, a recent study on chemically reduced negatively charged CdSe QDs reported a hot electron cooling time constant of 320 ps in the presence of reductant-induced electron traps, which may assist in cooling.[18] This suggests that the phonon-bottleneck limit for hot electron relaxation in CdSe QDs is slower than that of CdS, as predicted by theory.[23]

The slow cooling in CdS QD$^{(-)}$ (compared to neutral QDs) observed here is consistent with recent reports of reductive organic photocatalysis where negatively charged CdS QDs could drive transformations with reduction potentials more negative than the conduction band of the CdS QDs.[24, 25] For electrons to transfer to the molecular substrates, they would require excess energy above the conduction band edge. It has been hypothesized that such hot electrons are generated by trion recombination in negatively charged CdS QDs.[24, 25] Our work suggests that these hot electrons are not only energetic but also relatively *long lived*, which would allow electron transfer to organic substrates to outcompete electron cooling, enabling photocatalysis.

## Conclusions:

We have studied hot electron cooling in negatively charged CdS QDs and found this process to be relatively slow, occurring on a 100 ps timescale. Theoretical insights into the cooling mechanism indicate that this timescale is the phonon bottleneck limit for these particles, where the electronic cooling timescale can be fully captured by lattice phonon-assisted relaxation. By identifying the phonon bottleneck limit, we provide a quantitative target for designing QDs with optimally slow electron cooling and define the lower limit on the rates of hot electron harvesting processes. The fact that the bottleneck-limited cooling rate is relatively slow and the fact that we experimentally observe it in the negatively charged CdS QDs is promising for the application of CdS and similar QDs in light-harvesting applications.

## Methods:

### QD Synthesis

Synthetic conditions for the main batch of QDs studied in this work have been reported previously and the particle characterization is described there.[28] The solvent for all the measurements below was toluene because, as described in prior work,[28] it allowed for a high degree of charging over long times without sample degradation.

The QDs used for Figure S8 were synthesized using a previously reported procedure.[65] The cadmium oleate precursor was synthesized using a previously reported procedure.[66] The CdS QD synthesis was scaled up two-fold from the original Hamachi *et al* procedure.[65] In a 50 mL three neck bottom flask, 0.2443 g cadmium oleate , 0.2165 g oleic acid (99 %, Thermo Scientific Chemicals), and 20.1343 g n-hexadecane (≥ 99 % for synthesis, Sigma-Aldrich) were added and placed under Argon on a Schlenk line. The mixture was then heated to 60 °C and placed under vacuum for 1 hour. While under vacuum, the temperature was increased from 60 °C to 90°C in 10 °C increments. The solution was then placed under Argon and heated to 240 °C. After the solution reached 240 °C, a solution of 0.0399 g tetramethylthiourea (98%, Sigma-Aldrich) in 1.50 mL diphenyl ether (≥ 98 % for synthesis, Sigma-Aldrich) was injected. The reaction proceeded for 180 minutes after which the solution was cooled to 90 °C. The solution was then withdrawn from the flask and transferred into an Argon glovebox for purification without exposure to air. The QDs were precipitated with 40 mL acetone (≥ 99 %, Sigma-Aldrich) and centrifuged at 4400 rpm for 20 minutes. Supernatant was removed and QDs were dissolved in 5 mL hexanes (≥ 99 % anhydrous, Sigma-Aldrich). Unreacted cadmium oleate was removed by precipitation by addition of acetone in 0.5 mL increments up to a total of 6 mL. This solution was then centrifuged for 1 minute at 4400 rpm to separate the unreacted cadmium oleate from the colloidal QDs. The QDs in solution were moved to a new centrifuge tube and then precipitated with 40 mL of acetone. The solution was centrifuged at 4400 rpm for 20 minutes and the supernatant was discarded. The QDs were then washed with toluene/methanol (1 mL toluene (≥ 99.9 % anhydrous, Sigma-Aldrich)/2 mL methanol (≥ 99.9 % anhydrous, Sigma-Aldrich)) two additional times. The final oleate capped CdS QDs were dissolved in 500 μL of toluene (≥ 99 % anhydrous, Sigma-Aldrich) and stored in an Ar glovebox.

### UV-VIS Absorbance

Absorbance spectra were collected using a Cary 60 UV-vis spectrophotometer (Agilent). Samples were kept in a 1 cm x 1 cm cuvette sealed with a Kontes valve. Spectra were corrected for background solvent absorption. When used, a 405 nm continuous wave (CW) laser (Laserglow Technologies, LRD-0405-PFR) was directed at 90° angle relative to the direction of absorbance collection and the beam was about 1 cm below the absorbance collection beam.

### TA Spectroscopy

TA was collected using a setup that has been described previously.[67] Briefly, a regeneratively amplified Ti:Sapphire laser (Spectra-Physics, Solstice) produced ~100 fs pulses of 800 nm light at a repetition rate of 1 kHz. A portion was aligned into an optical parametric amplifier (Light Conversion, TOPAS-C) which was optimized to create a 365 nm pump pulse. We chose the pump wavelength of 365 nm because it is an isosbestic point between the absorbance of charged and neutral QDs so the total number of combined generated excitons (in neutral QDs) and trions (in charged QDs) is similar across different CW powers. The probe beam was derived from another portion of the 800 nm light by focusing into a 3 mm sapphire plate. Pump and probe beam overlap were optimized to achieve the strongest ΔA signal. The pump power was selected as to only create single excitations (excitons or trions); details are given in Section S2. Data was collected using a HELIOS spectrometer (Ultrafast Systems). Time points were sampled between negative 5 ps to 3.2 ns. Points before 0 ps were collected with 0.2 ps steps and points after 0 ps were collected using exponentially distributed steps starting with a 20 fs step with 500 total points each averaged for 1 s. 6 scans were completed for each condition tested. The QDs studied here had an OD of 0.4 at the 1S transition and were sealed in a custom made 2 mm quartz cuvette outfitted with a Kontes valve at the top to ensure a rigorously air-free environment.

To control the degree of charging in our ensemble of QDs, we introduced a 405 nm CW beam from a diode laser (Laserglow Technologies, LRD-0405-PFR) with a 1 mm beam diameter. We routed this beam to hit the cuvette approximately 1 cm away from the pump/probe overlap spot. The power of this CW laser was adjusted between 0 - 40 mW to control the populations of charged and neutral QDs. We note that the first charging and discharging cycle have different kinetics than the subsequent ones.[28] Similarly, we found that the excited state decay changes after the first charging cycle but remains constant after. For this reason, prior to TA measurements, we exposed the QDs to the 405 nm CW laser with 40 mW power for 1 hour to ensure that the measurements are in the regime where the excited state dynamics do not change in the absence of CW power (Figure S9). The sample stirring speed was kept consistent so that the laser power was the main factor which controlled the degree of photocharging.

**Theoretical Simulation**

A 4.5 nm diameter CdS QD ($Cd_{753}S_{753}$) was constructed by adding layers to a wurtzite seed with lattice parameter a = 4.17Å and c = 6.78Å. The structure was then relaxed using the Stillinger-Weber force field,[68] parameterized for II-VI nanostructures. We then applied the atomistic semi-empirical pseudopotential method to calculate the electronic structure of the relaxed QD. The pseudopotential consists of a combination of short- and long-range contributions,[23] which were fitted to replicate the band structure and electron-phonon coupling strength obtained from first-principles calculations in bulk CdS (see fitting parameters $a_0$–$a_4$ in Section S4). In particular, the long-range part of the potential was included to capture the electron-phonon coupling mechanisms mediated by the long-range Frölich-type interaction. The filter diagonalization technique[22] was applied to compute the quasi-electron energies $E_n$ and eigenstates $\psi_n(\mathbf{r})$ near the bottom of the conduction band, including 1S$_e$ and three near-degenerate 1P$_e$ states. The nuclear vibrations are described by a set of phonon frequencies $\omega_\alpha$ and coordinates $Q_\alpha$ which are determined by diagonalizing the Hessian matrix of the Stillinger-Weber potential at the equilibrium configuration. Figure 3d illustrates three vibrational modes that are strongly coupled to the electronic states of the QD. The electron-phonon couplings strength between state $\psi_n(\mathbf{r})$ and $\psi_m(\mathbf{r})$ mediated through mode $Q_\alpha$ is given by $V_{nm}^\alpha = \langle \psi_n | \frac{\partial v}{\partial Q_\alpha} | \psi_m \rangle$, where $v(\mathbf{r}; \mathbf{R})$ is the total pseudopotential.[21] The distribution of electron-phonon couplings among each mode can be represented by the spectral density defined as $J_{nm}(\omega) = \sum_\alpha \frac{(V_{nm}^\alpha)^2}{2\omega_\alpha} \delta(\omega - \omega_\alpha)$. Figure 3c (insert) plots the spectral densities for intrastate (1Se-1Se and 1Pe-1Pe) and interstate couplings (1Se-1Pe). These parameters from the above calculations were further used to parameterize a linear vibronic coupling (LVC) model Hamiltonian. The dynamics of hot electron relaxation were then described using a polaron-transformed master equation, which accounts for multi-phonon processes.[24,25]

## Associated Content:

Supporting Information Available: Early time dynamics of QDs, determination of pump power, spectral changes during charging/de-charging, theoretical methods, derivation of model, predicted ΔA signatures, comparison of fits to data, and comparison of kinetics before and after aging QDs.

## Corresponding Author

*Gordana Dukovic - Department of Chemistry and Renewable and Sustainable Energy Institute (RASEI), University of Colorado Boulder, Boulder, Colorado 80309, United States; Materials Science and Engineering, University of Colorado Boulder, Boulder, Colorado 80303, United States; orcid.org/0000-0001-5102-0958; Phone: +1-303-735-5297; Email: gordana.dukovic@colorado.edu

## Present Addresses

† Department of Chemistry, Brandeis University, 415 South Street, Waltham, Massachusetts, 02453, USA.

## Author Contributions

The manuscript was written through contributions of all authors. All authors have given approval to the final version of the manuscript.

## Acknowledgements

The experimental portion of this work was supported by the Air Force Office of Scientific Research under AFOSR Award Nos. FA9550-19-1-0083 and FA9550-22-1-0347. Simulations were supported by the U.S. Department of Energy, Office of Science, Office of Basic Energy Sciences, Materials Sciences and Engineering Division, under Contract No. DEAC02-05-CH11231 within the Fundamentals of Semiconductor Nanowire Program (KCPY23). Computational methods used in this work were developed with support from the U.S. Department of Energy, Office of Science, Office of Advanced Scientific Computing Research and Office of Basic Energy Sciences, Scientific Discovery through Advanced Computing (SciDAC) program, under award no. DE-SC0022088. Computational resources were provided in part by the National Energy Research Scientific Computing Center (NERSC), a U.S. Department of Energy Office of Science User Facility operated under Contract No. DEAC02-05CH11231. S.J.S. acknowledges support from the Edward L. King and Marion L. Sharrah fellowships at University of Colorado Boulder. K.E.S. acknowledges support from the National Science Foundation Grant No. CHE-2125978 awarded to the Research Corporation for Science Advancement for the Cottrell Fellowships Award No. 27980.

TOC Graphic:

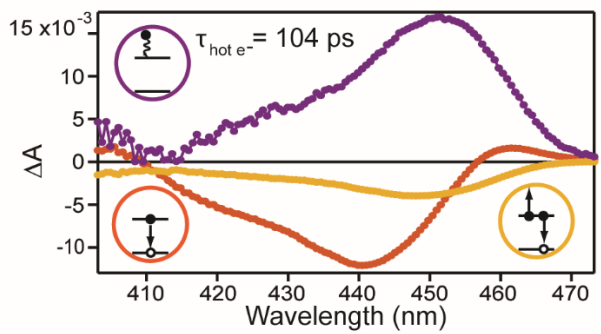

# Supporting Information for "Revealing the Phonon Bottleneck Limit in Negatively Charged CdS Quantum Dots"

*Skylar J. Sherman[1], Bokang Hou[2,3], Matthew J. Coley-O'Rourke[2], Katherine E. Shulenberger[1,†], Lauren M. Pellows,[1] Eran Rabani[2,3,4], Gordana Dukovic[1,5,6]\**

1. Department of Chemistry, University of Colorado Boulder, Boulder, Colorado 80309, United States

2. Department of Chemistry, University of California Berkeley, Berkeley, California 94720, United States

3. Materials Sciences Division, Lawrence Berkeley National Laboratory, Berkeley, California 94720, USA

4. The Raymond and Beverly Sackler Center of Computational Molecular and Materials Science, Tel Aviv University, Tel Aviv 69978, Israel

5. Department of Chemistry and Renewable and Sustainable Energy Institute (RASEI), University of Colorado Boulder, Boulder, Colorado 80309, United States

6. Materials Science and Engineering, University of Colorado Boulder, Boulder, Colorado 80303, United States

**Table of Contents:**

## Section S1: Initial Cooling of Hot Electrons in Neutral and Negatively Charged CdS QDs

To isolate the dynamics of hot electron cooling after initial pump excitation, the 1S bleach maximum intensity is monitored after resonantly pumping the 1S and the 1P transitions and the two decay traces are subtracted from each other.[1-3] When pumping resonantly with the 1S transition, the exciton (or trion if the QDs are charged) is excited into its lowest energy excited state. When the 1P transition is resonantly excited, the electron is hot, and the hole is cold. (In principle, the hole does have a very small difference in energy to the valence band edge, but this gap is quite small making it effectively cold). Because the 1S bleach maximum signal is dominated by electrons populating the conduction band edge in core-only cadmium chalcogenide QDs,[4,5] the rise of this signal reports on the hot electron cooling dynamics. Figure S1a shows the bleach maximum time traces at 441 nm for neutral QDs when resonantly pumped at the 1S and 1P transitions. When the 1S is excited, the rise of the bleach maximum is instantaneous and its apparent rise time of 143 fs is limited by the instrument response function of our TA setup. When the 1P is excited, the rate of hot electron cooling from the 1P to the 1S is slower than and convoluted with the instrument response function. The difference between the 1S and 1P pump reports on the hot electron cooling from the 1P to the 1S. Fitting the difference to a gaussian rise convoluted with an exponential decay results in a cooling time of 42 fs.

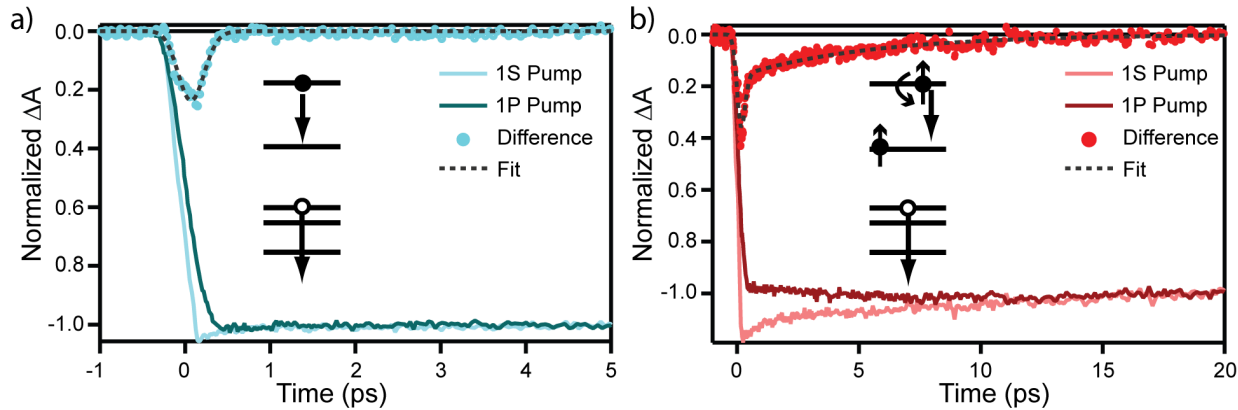

**Figure S1.** (a and b) Bleach maximum (441 nm) time-traces for 1S and 1P pump for neutral and charged QDs, respectively. Differences is shown as dots with fits shown as dashed lines.

Figure S1b shows the results of the same measurement for the sample illuminated with CW power of 40 mW, which contains both neutral and charged QDs. Compared to the neutral QDs (Figure S1a), the time traces for the 1S bleach maximum for the 1S and 1P resonant pumps take longer to converge. Similar behavior has previously been observed in negatively charged CdSe QDs and was attributed to a spin blockade.[6] In a negatively charged QD, the resident electron can either have a spin up or spin down. When this QD is excited, there is a 50/50 chance that the excited electron has the same spin as the resident electron. Due to the Pauli exclusion principle, if the excited electron has the same spin as the resident one, it (or the resident electron) must flip its spin for cooling to proceed, which slows down electron cooling. The difference between the 1S bleach signal between the 1S and 1P pump is fit well to a gaussian rise convolved with two exponentials. The first exponential has the lifetime of 50 fs, which corresponds to neutral QDs and charged QDs with opposite electron spins. The second exponential has a time constant of 5 ps, corresponding to hot electron cooling in particles that require a spin flip. This is very similar to the time constant found for a spin flip in CdSe QDs of 9 ps.[6]

## Section S2. Determination of Pump Power

We performed the transient absorption (TA) experiments at low enough fluence to create either excitons or trions and avoid multiexcitons. To determine this power, we collected ΔA spectra of neutral CdS QDs at 3 ns as a function of pump power (Figure S2). Because neutral QDs could in principle become negatively charged from pump illumination especially at high powers, we collected ΔA spectra with short collection times at only one pump-probe delay (3 ns), and we allowed the particles to stir in the dark for 15 minutes before each measurement. At this delay time, all excitations beyond single excitons (ie, biexcitons, triexcitons, etc.) will have decayed into single excitons.[7-9] Since absorption at wavelengths well above the band edge obeys Poisson statistics,[7] the late time signal is related to pump power by Equation (S1):

$$\Delta A(3ns) = A * (1 - exp(-C * J)), \quad (S1)$$

where A is an amplitude, C is a constant that relates the average number excitations to the power by $\langle N \rangle_{exc} = C * J$ and is related to the pump/probe overlap, sample concentration, and absorption cross section and J is the power. By fitting the ΔA at 3 ns to Equation (S1) we can relate the pump power to the average number of excitations (Figure S2). For the TA measurements, we choose a power where <N>=0.15. At this value of <N>, only 7% of excitations result in biexcitons or higher.

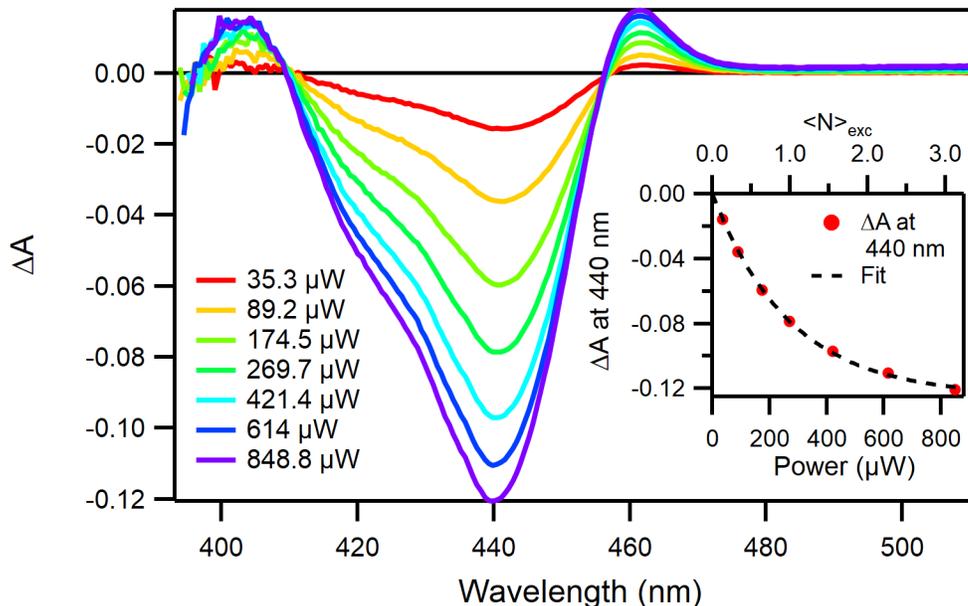

**Figure S2.** ΔA at spectra at 3 ns as a function of pump power. Inset shows the ΔA at 440 nm as a function of power with fit to Equation S1 shown as dashed black line.

## Section S3: Changes to TA Spectra During Charging and De-charging

To observe the changes in the TA spectra due to charging and discharging, we collected the ΔA spectra at 10 ps while the equilibrium between charged and neutral QDs is established (Figure S3). We found that the reduction of the ΔA strength occurred over the course of minutes, the timescale of charging observed in steady-state experiments.[10] The ΔA spectrum recovers to the match the spectrum before illumination with 405 nm CW laser because charging is reversible.

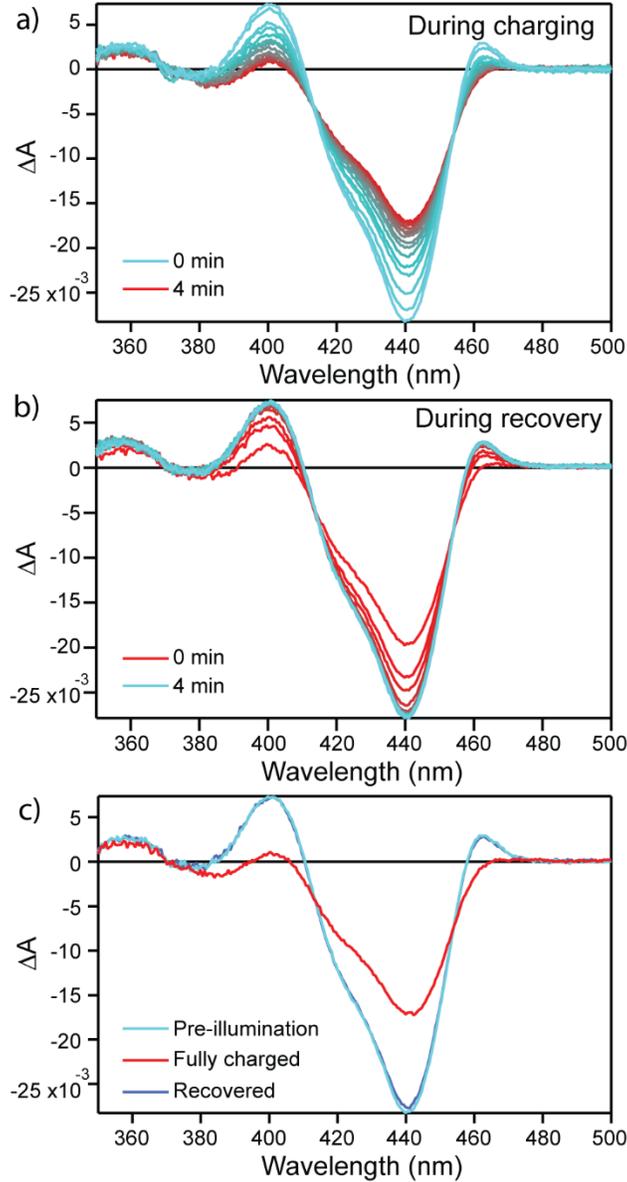

**Figure S3.** (a) ΔA spectrum collected at 10 ps as a function of time after illumination with a 40 mW 405 nm CW laser begins. (b) Same as for (a) but time after the 405 nm CW laser is blocked. (c) Comparison of ΔA spectrum before illumination with 405 nm CW laser, at maximum charging, and after recovery.

## Section S4: Semi-Empirical Pseudopotentials with Long-Range Effects

For computation of the electronic structure of the CdS QD, we employed a semi-empirical pseudopotential method, which has been well-tested and verified by many authors.[11-13] The total potential is decomposed as a sum of atom-centered functions, which are expressed in reciprocal space as,[14]

$$V_{tot}(\mathbf{g}_i, \mathbf{g}_j) = \sum_\mu \left[ a_0^\mu \frac{|\mathbf{g}_i-\mathbf{g}_j|^2 - a_1^\mu}{a_2^\mu \exp(a_3^\mu |\mathbf{g}_i-\mathbf{g}_j|^2) - 1} - 4\pi a_4^\mu \frac{\exp(-|\mathbf{g}_i-\mathbf{g}_j|^2/(4\lambda^2))}{|\mathbf{g}_i-\mathbf{g}_j|^2} \right] e^{-i(\mathbf{g}_i-\mathbf{g}_j)\cdot \mathbf{R}_\mu}. \quad (S2)$$

Here, $\mathbf{g}_i$ labels the reciprocal space basis (plane waves), $\mathbf{R}_\mu$ denotes the position of atom $\mu$, and $\{a_0, a_1, a_2, a_3, \lambda\}$ are parameters that are tuned to reproduce the bulk band structure and electron-phonon coupling computed from *ab initio* theory. Equation S2 includes a well-established short-range function in real space,[12, 13, 15] and a long-range term which is the Fourier transform of the Coulomb potential $a_4^\mu \mathrm{erf}(\lambda|\mathbf{r}-\mathbf{R}_\mu|)/|\mathbf{r}-\mathbf{R}_\mu|$ in real space.[14] This enables our model to capture various important electron-phonon coupling mechanisms,[14] such as the long range Frölich-type interaction. To remove surface trap states from the gap and cancel macroscopic electric fields generated by the faceted surfaces of the quantum dot, we attach fictitious ligand potentials to the dangling bonds at the surface.[14]

|    | $a_0$      | $a_1$    | $a_2$       | $a_3$     | $a_4$   | $\lambda$ |
|----|------------|----------|-------------|-----------|---------|-----------|
| Cd | -31.450808 | 1.665222 | -0.16198998 | 1.672854  | -0.625  | 0.2       |
| S  | 7.665165   | 4.444229 | 1.384734    | 0.2584866 | 0.625   | 0.2       |

**Table S1:** Pseudopotential coefficients $a_0$ to $a_4$ and $\lambda$ for Cd and S.

### Section S5: Model Hamiltonian and Hot Electron Relaxation Rates

We used the semi-empirical pseudopotentials discussed above to parameterize the model Hamiltonian $H = H_{el} + H_{nu} + H_{el-nu}$ that describes the interaction between electrons and nuclear vibrations[11, 14]

$$H_{el} = \sum_n E_n |\psi_n\rangle\langle\psi_n| \quad (S3)$$

$$H_{nu} = \sum_\alpha \tfrac{1}{2} P_\alpha^2 + \tfrac{1}{2} \omega_\alpha^2 Q_\alpha^2 \quad (S4)$$

$$H_{el-nu} = \sum_{n,m} |\psi_n\rangle\langle\psi_m| \sum_\alpha V_{nm}^\alpha Q_\alpha \quad (S5)$$

For the electronic part, $E_n$ and $\psi_n$ are the eigenenergy and eigenstates of the quasi-electron Hamiltonian $h_e$, which is parameterized by semi-empirical pseudopotentials

$$h_e = \frac{1}{2}\nabla^2 + \sum_\mu v_\mu(|\mathbf{r} - \mathbf{R}_\mu^0|). \tag{S6}$$

The filter diagonalization technique[16] was applied to calculate the quasi-electron states near the bottom of the conduction band. The nuclear degrees of freedom were represented using the Stillinger-Weber force field,[17] with the reference geometry $\mathbf{R}^0$ being the energy minimum of the Stillinger-Weber potential energy surface. Under the harmonic approximation, the normal mode frequencies $\omega_\alpha$ and coordinates can be calculated by diagonalizing the Hessian matrix at $\mathbf{R}^0$. $P_\alpha$ and $Q_\alpha$ denote the mass-weighted normal mode momentum and coordinates, respectively, for mode $\alpha$. The transformation between atomic and normal modes is defined as $Q_\alpha = \sum_{\mu,k} \sqrt{m_\mu} E_{\mu k,\alpha}(R_{\mu k} - R_{\mu k}^0)$, where $k = x, y, z$ and $m_\mu$ is the mass of atom $\mu$. The electron-phonon coupling between the state $|\psi_n\rangle$ and $|\psi_m\rangle$ with respect to mode $\alpha$ is denoted as $V_{nm}^\alpha$, which could be related to the derivative of the pseudopotential by[18]

$$V_{nm}^\alpha \equiv \langle\psi_n|\frac{\partial h_e}{\partial Q_\alpha}|\psi_m\rangle = \sum_{\mu,k} \frac{1}{\sqrt{m_\mu}} E_{\mu k,\alpha} \langle\psi_n|\frac{\partial v_\mu}{\partial R_{\mu k}}|\psi_m\rangle. \tag{S7}$$

The transition rate $k_{n\to m}$ from state $|\psi_n\rangle$ to $|\psi_m\rangle$ (or 1P to 1S transition in this case) was calculated through Fermi's golden rule based on the polaron transformed model Hamiltonian. The purpose of performing the polaron transform is to renormalize the interstate couplings $V_{12}^\alpha$, improving the performance of perturbation theory while incorporating multi-phonon processes.[19] The off-diagonal term $g_{nm} \equiv e^{S_n} \sum_\alpha V_{nm}^\alpha Q_\alpha e^{-S_m}$, with $S_n = -\frac{i}{\hbar}\sum_\alpha \frac{V_{nn}^\alpha}{\omega_\alpha^2} P_\alpha$ was treated as perturbation, and based on the time-dependent form of Fermi's golden rule, the transition rate between states $|\psi_n\rangle$ and $|\psi_m\rangle$ can be written as[20]

$$k_{n\to m}(t) = \frac{1}{\hbar^2}\int_{-t}^{t} d\tau e^{i\frac{\varepsilon_n - \varepsilon_m}{\hbar}\tau}\langle g_{nm}(\tau)g_{mn}(0)\rangle_{eq}, \tag{S8}$$

where $\langle \cdot \rangle_{eq}$ represents the average over nuclear degrees of freedom in thermal equilibrium. The correlation function can then be derived quantum-mechanically.[18] Given the expression for the golden rule rates $k_{n \to m}(t)$, we can formulate a master equation for hot electron relaxation through $N$ electronic states. The population $p_n(t)$ at state $\psi_n$ satisfies

$$\frac{dp_n}{dt} = \sum_{\substack{m \\ m \neq n}} p_m k_{m \to n}(t) - \sum_{\substack{m \\ m \neq n}} p_n k_{n \to m}(t). \tag{S9}$$

Main Text Figure 3 plots the population $p_n(t)$ for electron relaxation among three 1P and 1S states. The initial population was placed in the highest energy 1P state.

## Section S6: Error Estimation of Hot Electron Relaxation Rates

We found that the relaxation rate for the hot electron is very sensitive to the 1P-1S gap, therefore, an error analysis is important when comparing the theoretical calculations with the experiment. If we assume the uncertainty of the QD size measurement is about ± 0.5 nm, this will cause a change of energy gap of ± 30 meV based on the pseudopotential calculations. Figure S3 shows the 1P-1S transition rate as a function of the energy gap, which is calculated through the Fourier transform of the time correlation function $\langle g_{nm}(\tau) g_{mn}(0) \rangle_{eq}$

$$k_{n \to m}(\Delta \varepsilon_{nm}) = \frac{1}{\hbar^2} \int_{-\infty}^{\infty} d\tau e^{i \frac{\Delta \varepsilon_{nm}}{\hbar} \tau} \langle g_{nm}(\tau) g_{mn}(0) \rangle_{eq}. \tag{S10}$$

The relaxation rate scales exponentially with the energy gap, which gives an estimate of the uncertainty for the rate as 0.043 [-0.036, +0.41] ps$^{-1}$, or in terms of time constant: 23.0 [-20.8, + 121] ps. The experimental measurement is within this uncertainty interval.

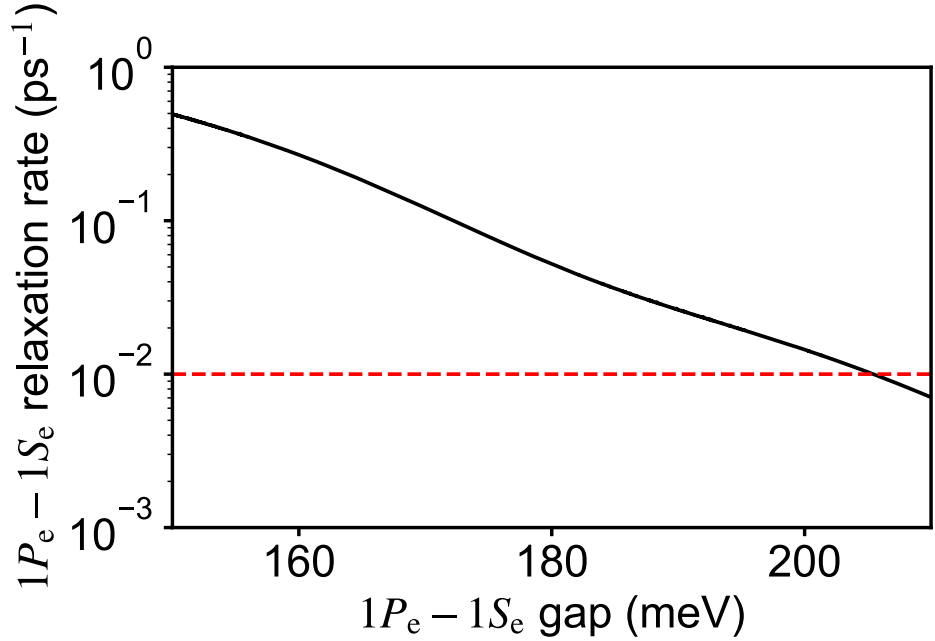

**Figure S4.** Hot electron relaxation rate as a function of the 1S-1P gap (solid black). The rate approximately scales exponentially as a function of the energy gap. The experimentally measured rate is shown as the red dashed line.

## Section S7: Derivation of the Kinetic Model for Electron Decay After 30 ps

We first define the first order rate equations for the processes shown schematically in Main Text Figure 4 and Figure S5.

$$\frac{dN_x}{dt} = -k_x N_x \tag{S19}$$

$$\frac{dN_t}{dt} = -k_t N_t \tag{S20}$$

$$\frac{dN_h}{dt} = k_t N_t - k_h N_h \tag{S30}$$

$N_x$, $N_t$, and $N_h$ are the time-dependent populations of excitons, trions, and hot electrons, respectively. $k_x$, $k_t$, and $k_h$ are the rate constants for exciton, trion, and hot electron decay, respectively. We assume there is no initial population of hot electrons and that there is only an initial population of excitons and trions, $N_x(t=0) = N_{x0}$ and $N_t(t=0) = N_{t0}$, respectively. The solutions to the coupled set of equations above with the imposed initial conditions give:

$$N_x = N_{x0}e^{-k_x t} \tag{S31}$$

$$N_t = N_{t0}e^{-k_t t} \tag{S32}$$

$$N_h = \frac{N_{t0}k_t}{k_h - k_t}\left(e^{-k_t t} - e^{-k_h t}\right) \tag{S33}$$

We then assume that each species has a unique spectroscopic signature and the observed ΔA will be the sum of the signals of each species weighted by their respective time-dependent populations.[8]

$$\Delta A = \Delta A_x N_x + \Delta A_t N_t + \Delta A_h N_h \tag{S34}$$

Substituting in equations S31, S32, and S33 into equation S34 gives:

$$\Delta A = \Delta A_x N_{x0} e^{-k_x t} + \Delta A_t N_{t0} e^{-k_t t} + \Delta A_h \frac{N_{t0} k_t}{k_h - k_t}\left(e^{-k_t t} - e^{-k_h t}\right) \tag{S35}$$

Rearrangement leads to the **Main Text Equation 1**:

$$\Delta A = \Delta A_x N_{x0} e^{-k_x t} + N_{t0} e^{-k_t t}\left(\Delta A_t + \Delta A_h \frac{k_t}{k_h - k_t}\right) \tag{S36}$$

$$- \Delta A_h \frac{k_t}{k_h - k_t} e^{-k_h t}$$

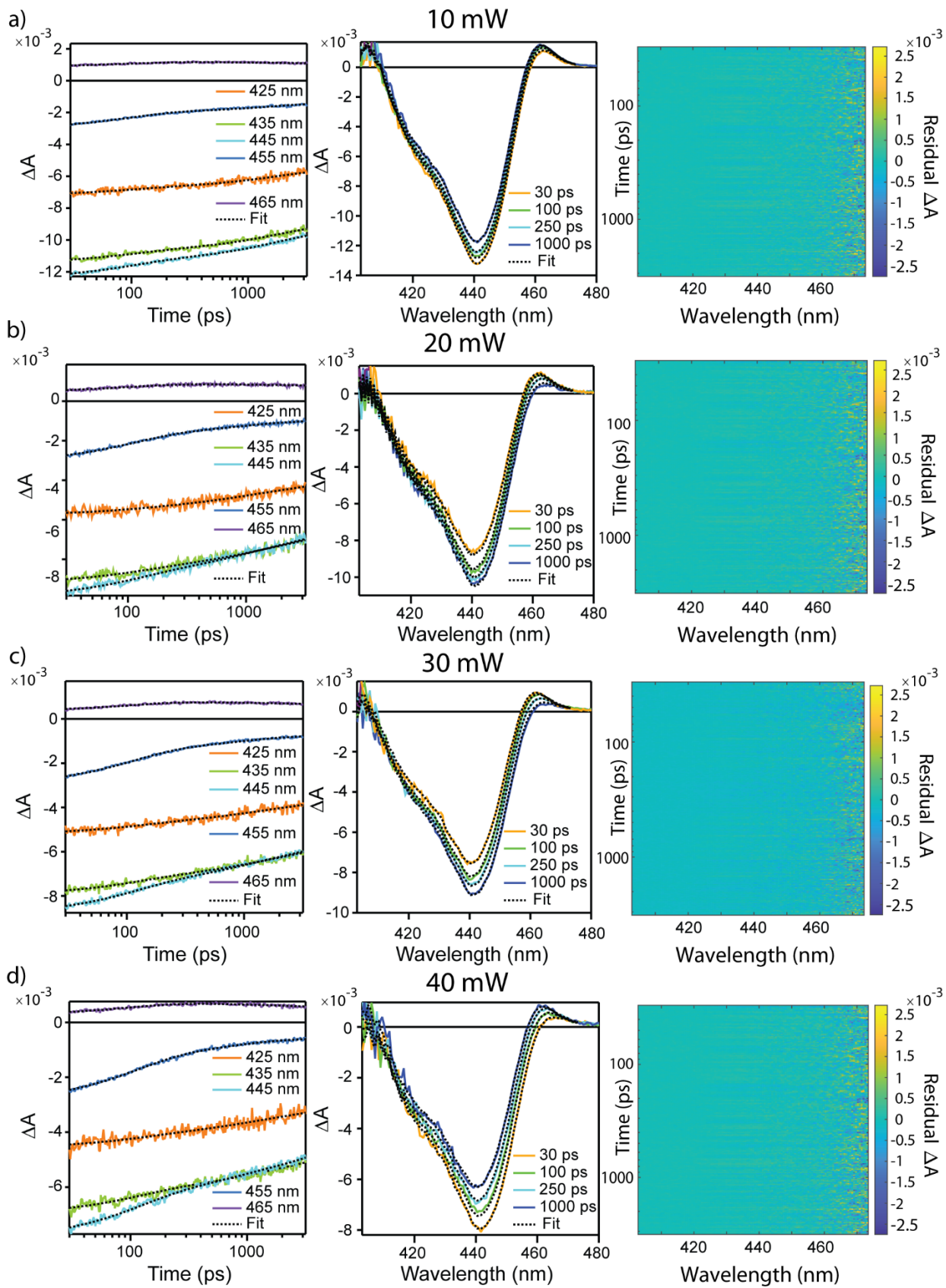

**Figure S5.** (a-d) Comparison of fit to time-traces from 30 ps to 3.2 ns at select wavelengths (left), comparison of fit to spectra at select time points (middle), and residual plot showing error from fit at all fit wavelengths and times (right) for 0, 10, 20, 30 and 40 mW CW powers.

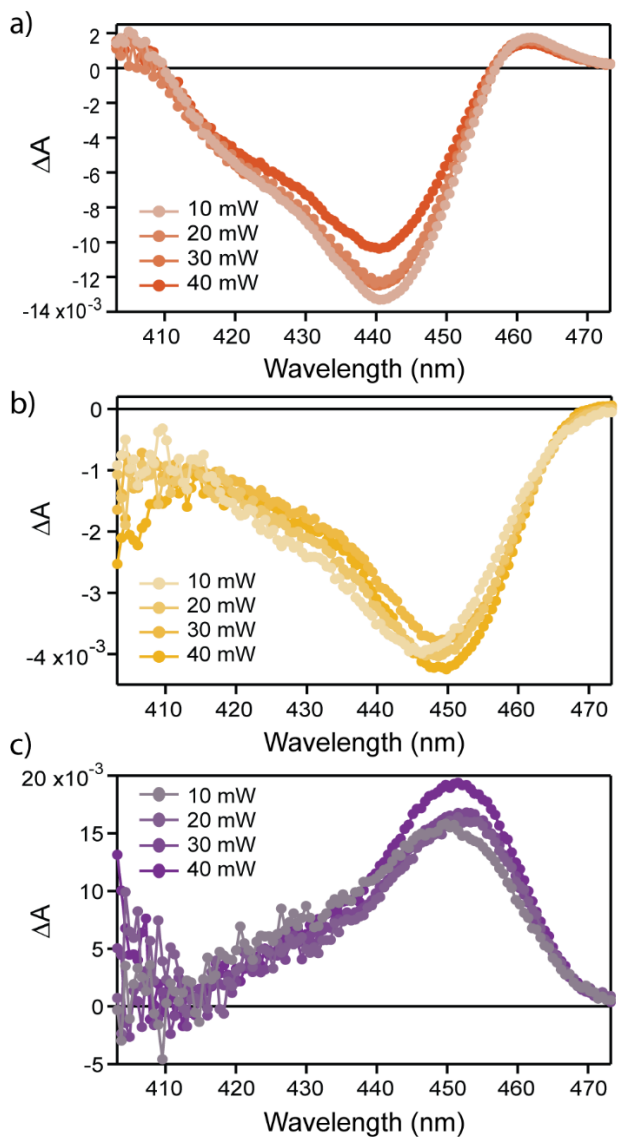

**Figure S6.** Extracted spectra for (a) $\Delta A_x$, (b) $\Delta A_t$, and (c) $\Delta A_h$ for each CW power.

| Power (mW) | 10 | 20 | 30 | 40 |
|---|---|---|---|---|
| Nx | 0.88 | 0.70 | 0.60 | 0.60 |
| Ntx | 0.48 | 0.62 | 0.67 | 0.64 |
| Total | 1.36 | 1.32 | 1.27 | 1.24 |
| $\chi_x$ | 0.65 | 0.53 | 0.47 | 0.48 |
| $\chi_{tx}$ | 0.35 | 0.47 | 0.53 | 0.52 |

**Table S3.** Fitted number of excitons, trions and total excitations for each CW power studied. Also given is the relative fraction of excitons and trions, $\chi_x$ and $\chi_{tx}$, respectively.

## Section S8: Discussion of Trion Signature ΔA Strength

As seen in Figure S5d and S5f if the steady-state absorbance is bleached by 50% upon charging, as is commonly assumed, the trion ΔA spectrum would be only slightly weaker than the exciton ΔA. This is clearly not the case in our measurements, as seen in the experimental data in Figure S3 and Main Text Figure 2b, where we see that increasing degree of photocharging leads to a decreasing bleach intensity in the TA spectra (the timescale is too short for there to be appreciably different band edge electron decay). Furthermore, in the extracted spectra in Main Text Figure 4, the maximum trion ΔA amplitude is only 33% of the maximum amplitude of the exciton ΔA. The likely explanation for this discrepancy is that an electron at the conduction band edge bleaches the 1S transition by more than 50% (similar to what is observed in photoexcited QDs).[8, 9, 21] To illustrate this, we write the experimentally observed ratio in terms of the respective transitions involved in the observed ΔA spectrum: $\frac{\Delta A_x}{\Delta A_{x^-}} = [-(G \to X) + (X \to XX)]/-(G^- \to X^-)$ where $\frac{\Delta A_x}{\Delta A_{x^-}}$ is the ratio of the exciton to trion ΔA at their respective maximal magnitudes, $(G \to X)$ is the strength of the ground to exciton transition, $(X \to XX)$ is the strength of the exciton to biexciton transition and $(G^- \to X^-)$ is the strength of the ground charged state to trion transition. We then define $\alpha$ as the fraction of the steady state absorbance that is bleached once a QD becomes charged. This leads to: $\frac{\Delta A_x}{\Delta A_{x^-}} = \frac{1}{\alpha} - \frac{(X \to XX)}{\alpha(G \to X)}$. While we do not know the $\frac{(X \to XX)}{(G \to X)}$ ratio precisely, we impose a lower bound of 1.3 and upper bound of 2, which spans a range that is larger than what has been experimentally determined for CdSe QDs.[8, 9] By inputting the experimentally determined $\frac{\Delta A_x}{\Delta A_{x^-}}$ ratio and these bounds, we arrive at $0.17 < \alpha < 0.26$. This corresponds to a steady state absorbance bleach of between 74 and 83% when an electron is added to the conduction band edge. Due to this uncertainty, we do not report the degree of charging in terms of number of electrons occupying the conduction band edge.

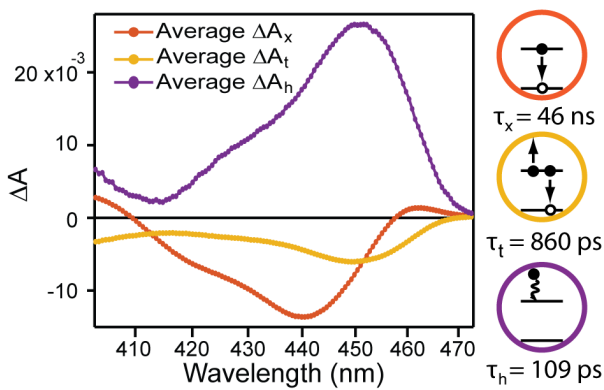

**Figure S7.** Component spectra for exciton, trion and hot electron after illumination with 405 nm CW laser with average power of 13.4 mW for 8.3 hours. Decay time constants with associated species are shown to the right. Extracted time constants are very similar to those found before illumination shown in Figure 5.

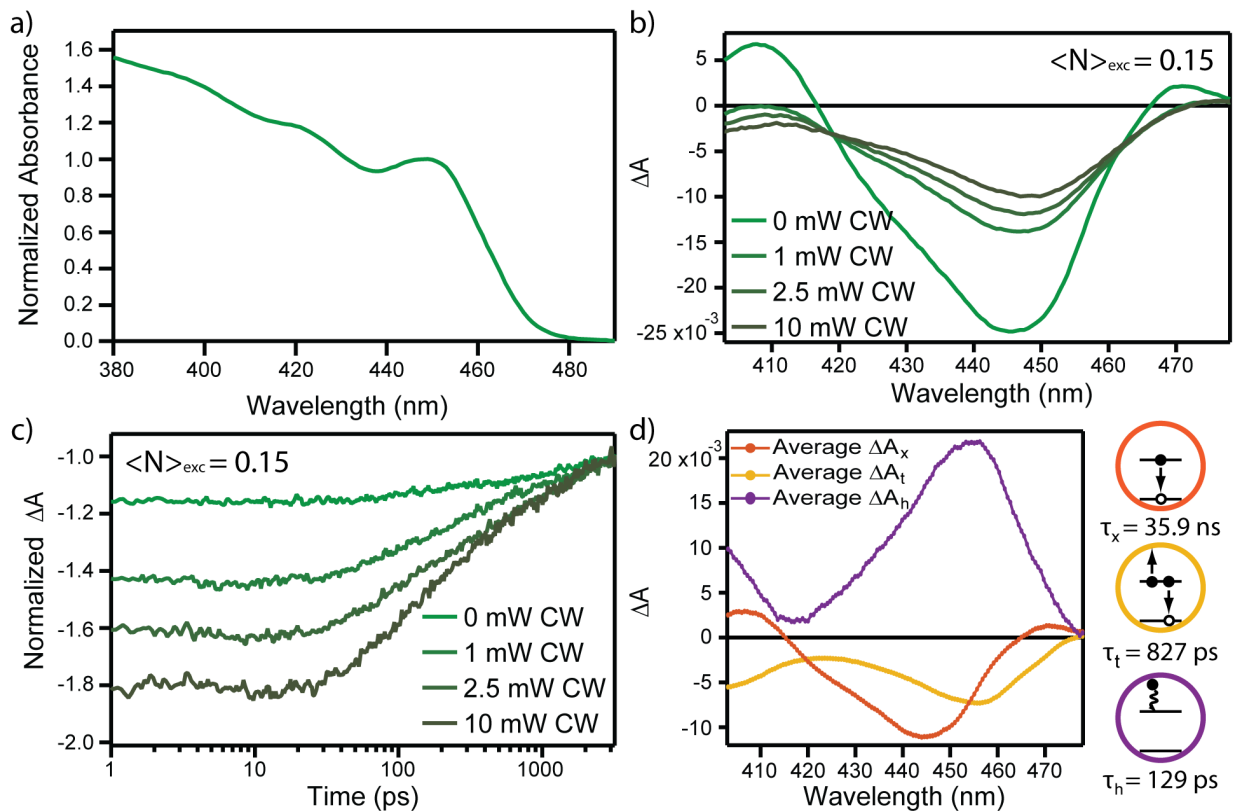

**Figure S8.** (a) Absorption spectrum for QDs with an average diameter of 4.6 nm. (b) ΔA spectrum at 20 ps as a function of 405 nm CW power. (c) ΔA time trace at 446 nm as a function of 405 nm CW power. (d) Fitted spectral signatures for the exciton, trion and hot electron along with time constants of decay shown to the right.

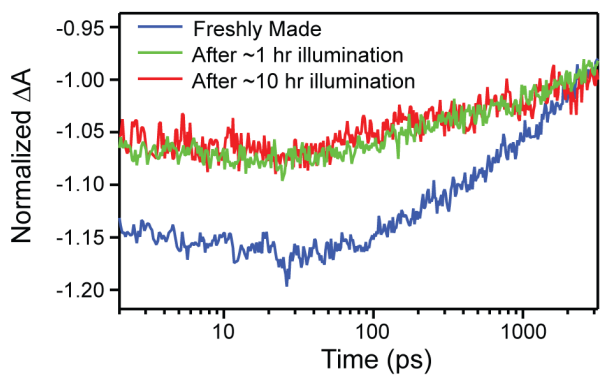

**Figure S9.** Comparison of normalized time-traces at the bleach maximum (441 nm) between freshly made CdS QDs and after illumination with the CW laser. An initial change in kinetics is observed but stabilizes after less than an hour of illumination.

**Supporting References:**